\documentclass[aps,prl,reprint,twocolumn]{revtex4-1}
\usepackage{amssymb}
\usepackage{epsfig}
\usepackage{graphicx}
\usepackage{amsmath}
\usepackage{subfigure}
\usepackage{array,color}
\usepackage{dcolumn}
\usepackage{bm}

\usepackage{amssymb}
\usepackage{epsfig}
\usepackage{graphicx}
\usepackage{amsmath}
\usepackage{subfigure}
\usepackage{array,color}
\usepackage{dcolumn}
\usepackage{bm}
\begin{document}

\title{Majorana Edge Modes of Kitaev Chain with Multiple Time Periodic Driving}
\author{Huan-Yu Wang$^{1,2}$, Lin Zhuang$^3$, and W.M. Liu$^{1,2,4}$ }
\email{wliu@iphy.ac.cn}
\address{$^1$Beijing National Laboratory for Condensed Matter Physics, Institute of Physics, Chinese Academy of Sciences, Beijing 100190, China}
\address{$^2$School of Physical Sciences, University of Chinese Academy of Sciences, Beijing 100190, China}
\address{$^3$State Key Laboratory of Optoelectronic Materials and Technologies, School of Physics, Sun Yat-Sen University, Guangzhou 510275, China}
\address{$^4$Songshan Lake Materials Laboratory, Dongguan, Guangdong 523808, China}
\begin{abstract}
Floquet Majorana edge modes capture the topological features of periodically driven superconductors.  We present a Kitaev chain with multiple time periodic driving and demonstrate  how the avoidance of band crossing is altered, which gives rise to new regions supporting Majorana edge modes. A one dimensional generalized method was proposed to predict Majorana edge modes via the Zak phase of the Floquet bands. We also study the time independent effective Hamiltonian at high frequency limit and introduce diverse
index to characterize topological phases with different relative phase between the multiple driving. Our work enriches the physics of driven system and paves the way for locating Majorana edge modes in larger parameter space.
\end{abstract}

\pacs{03.75.-b, 05.45.Yv, 34.50.-s}

\maketitle

Topolgical state of matter is an intriguing topic, and has been  studied intensively for years. Topological superconductivity is one of the most attractive theme in this subject for its topological excitation, Majorana fermion~\cite{Read-prb-2000,Ryu-prl-2002, Xiaoliang-PRL-2009, Cheng-prl-2009, Akhmerov-prl-2011, Eran-prb-2011, Xiaoliang-RMP-2011, Sumanta-prl-2012, Doru-prl-2012, Satoshi-prl-2012, Abolhassan-prb-2013, Sato-rep-2017, Ziesen-prb-2019,xiaoyu-prl-2019, Shingo-prl-2019}, which is its own antiparticle. The non-abelian statistics~\cite{Sh-prl-2011, Sau-prb-2011, Netanel-prx-2012, David-Natcomm-2013} of Majorana edge modes empower it as a novel prospect for quantum computation~\cite{Alicea-Natphy- 2011, liu-prl-2012}. Thus, methods on how to generate Majorana edge modes are gaining increasing attention~\cite{Fu-PRL-2008, Sau-PRL-2010, TDStanescu-prb-2011,Nadj-Perge-science-2014, Steven-RMP-2015, He-scince-2017}. Conventional techniques of creating  Majorana edge modes are via unpaired spin polarized fermions in quantum nanowires~\cite{Kitaev-Phys-Usp-2001,Bolech-prl-2007,Mourik-science-2012,Anindya-Natcomm-2012,Randeep-prb-2019} or superconductors in proximity with topological insulators or semiconductors~\cite{Hsuan-prl-2018,Bommer-prl-2019}. However, topological nontrivial regions with Majorana edge modes gained above are quite limited in parameter space.

In recent years, Floquet engineering emerges as a new protocol for designing topological states of matter~\cite{pz-prl-2011, Hubener-Natcomm-2017, qing-prl-2019}, which correspondingly brings  the concept Floquet Majorana edge modes. The main idea of Floquet engineering lies in driving the physical parameter periodically with time. In contrast to the adiabatic limit, where the system remains in the eigenstate at each instantaneous time,  driven system may absorb quantized energy from external fields, featuring non-equilibrium properties. Hence, the Floquet version of system may exhibit fruitful topological properties~\cite{Leon-prl-2010, Rudner-prb-2010, Rudner-prx-2013, Grushin-prl-2014, Erhai-prl-2014, Titum-prx-2015, Luca-Natcomm-2015, Flaschner-science-2016, Potter-prx-2016, Erhai-pra-2017, Roy-prb-2017,Eckardt-prl-2019}. For a driven Kitaev chain, results have shown that Majorana edge modes can be sustained within a larger parameter space~\cite{Benito-prb-2014,Daniel-prb-2017, Zeng-prb-2017,  Daniel-prl-2018}. Besides, multiple Majorana edge modes are discovered with low driving frequency~\cite{Manisha-prb-2013}. However, all results above are obtained within a frame of single time periodic driving. A natural question appears to us, how will the multiple time periodic driving affect the Majorana edge modes?

In this Letter,  we demonstrate that due to the avoidance of bands crossing, multiple time periodic driving are capable of generating gaps at different regions, leading to extraordinary different topological phase transitions, and larger parameter spaces with Majorana edge modes. To identify the Majorana edge modes, we propose a one dimensional method based on the Zak phase of Floquet bands, which can also be generalized to other 1D driven systems. Furthermore, given no degeneracy  in phase bands at high frequency limit, time independent effective Hamiltonian is utilized. Regardless of the relative phase $\phi$ between the multiple driving, our system always possesses particle-hole symmetry, which is the prerequisite for the existence of  Majorana edge modes. Besides, the relative phase determines the index to classify different topological phases. When $\phi\neq n\pi, n\in Z$, differences between single and multiple time periodic driving are manifested, where the winding number has to be replaced by the Zak phase.

We consider a spin polarized  Kitaev chain with multiple time periodic driving in the thermodynamic limit $N\gg1$, the  Hamiltonian is  of the form:
\begin{equation}\label{eq:hamilton}
\begin{aligned}
H(t)\!=&-r(t)\sum_{i=1}^{N}(c_{i+1}^{\dagger}c_{i}+H.c.)-\mu_0\sum_{i=1}^{N}c_i^{\dagger}c_{i}\\
&-\Delta(t)\sum_{i=1}^{N}(c_i^{\dagger}c_{i+1}^{\dagger} +H.c.),
\end{aligned}
\end{equation}
where $c_i^{\dagger}$ represents creating a fermion at site $i$. $r(t)=r_0+r_1\cos(\omega t)$ denotes the hopping with harmonic driving, $\Delta(t)=\Delta_0+\Delta_1\cos(\omega t+\phi)$ is the time periodic pairing between nearest sites. $\phi$ describes the relative phase between the multiple driving.
Hence, the Hamiltonian possesses the periodicity, $T=2\pi/\omega$, and Floquet theory can be applied. Since there are no well defined ground states in driven system, the topological properties are typically characterized by the time evolving operator:
\begin{equation}
U(T)\!=\!\textit{T}e^{-i\int_0^{T}H(t)dt}\!=\!e^{-i\epsilon T},
\end{equation}
which is treated the same way as the Hamiltonian in the static case. To identify different topological regions, we focus on the Floquet edge modes.  Different from the static case, driven Kitaev chain not only possesses Majorana zero edge modes (MZMs) but also anomalous edge modes located at $\pm\frac{\pi}{T}$, Majorana $\pi$ modes (MPMs) , and such edge modes can not be shifted to the bulk by expanding quasi energy $\epsilon\!=\!\epsilon+p\omega$, which gives its robustness. Thus, different topological phases can be classified via the number of MZMs and MPMs. The driven Kitaev chain should be indexed by $Z \times Z$. To get an overall comprehension, we set $\phi=\frac{\pi}{2}$ at first. Figs. 1(a), (b) depict the  Kitaev chain with single time periodic  pairing, hopping. It is manifested that, unlike the static case, the system still remains topological nontrivial for $\mu>2r_0$ with MPMs.
\begin{figure}[t]
 \centering
\includegraphics[width=8.5cm,height=8.5cm]{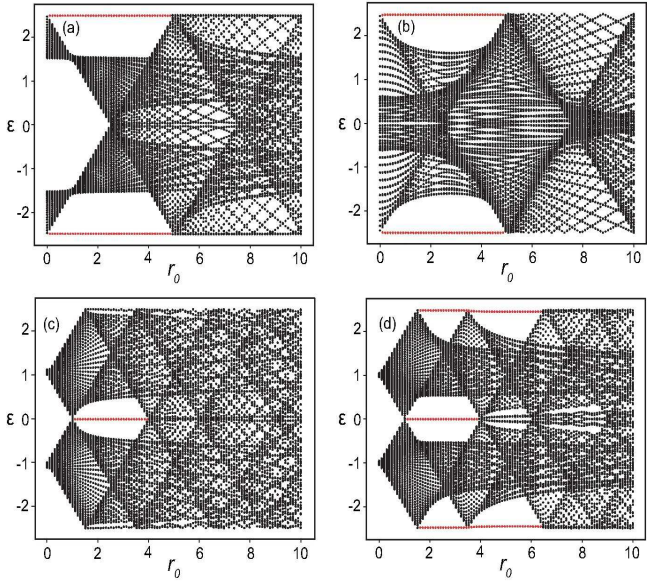}
\caption{\label{fig:floquet spectrum}(Color online) The Floquet spectrum as a function of $r_0$ for a driven Kitaev chain of 100 sites and driving frequency $\omega=5$.  MZMs and MPMs are marked with red dashed line. (a) MPMs in Kitaev chain with time periodic pairing, where $ \Delta_1=2, \Delta_0=0.5, \mu=5$.  (b) MPMs in Kitaev chain with time periodic hopping, where $r_1=2, \Delta_0=5, \mu=5$. (c) MZMs in Kitaev chain with time periodic hopping, where $r_1=2, \Delta_0=0.5, \mu=2$. (d) MPMs and MZMs in Kitaev chain with both time periodic pairing and hopping, where $r_1=2,\Delta_1=2, \Delta_0=0.5, \mu=2$.}
\end{figure}
To clarify the differences between single and multiple time periodic driving, we turn on the time periodic pairing and hopping sequently. Fig. 1(c) remarks that for a single time periodic hopping, there are no MPMs with small pairing amplitude~\cite{see supplemetary}. Need to mention that although there are edge modes offset from $\pm\frac{\pi}{T}$, such edge modes are fermion modes and can be shifted to bulk by expanding quasi energy. Therefore, it is still of topological trivial cases. Comparing  Fig. 1(c) with Fig. 1(d) manifests that, MPMs can be recovered by turning on the time periodic pairing.

To explain above phenomenons, we turn to the Floquet Hamiltonian $[H-i\partial_t]\Phi(t)\!\!=\!\!\epsilon\Phi(t), \Phi(t)\!\!=\!\!\Phi(t+T)$. With the Fourier transformation: $\Phi(t)\!\!=\!\!\sum_{m=-\infty}^{+\infty}\psi(m)e^{im\omega t}$, we obtain the Floquet Hamiltonian in frequency space $\tilde{H}_{m,m'}\!=\!\frac{1}{T}\int_{0}^{T} dt H(t)e^{i(m-m')\omega t}+m'\omega \delta_{m,m'}$. The static part of the driven Kitaev chain are transformed to $\tilde{H}_{m=m'}\!=\!-(r_0\cos(k)\!+\!\frac{\mu}{2})\sigma_z\!+\!\Delta_0\sin(k)\sigma_y\!+\!m\omega I$. The time periodic hopping and pairing contributes to off diagonal blocks in Floquet Hamiltonian respectively as: $\tilde{H}_{m'=m\pm 1}^{hopping}\!\!=\!\!\mp\frac{ r_1}{2}\cos(k)\sigma_z$, $\tilde{H}_{m'=m\pm 1}^{pairing}\!\!=\!\!\mp i\Delta_1\sin(k)\sigma_y$. Both driving are capable of generating gaps due to the avoidance of bands crossing. For example, since the first quasi energy Brillouin zone is localized in m, we truncate the Floquet Hamiltonian at $m=3$ as an approximation. Fig. 2 represents the Floquet quasi energy in frequency space.  By comparing Figs. 2(a),(b), it is exhibited that the  time periodic pairing introduces gaps at $\pm\frac{\omega}{2}$, which gives birth to MPMs.

Numerical calculations reveal that with varying $r_0$,  quasi energy (frequency space) $\pm\frac{\omega}{2}, 0$ gaps close and reopen at exactly the place where MPMs, MZMs merge. To confirm topological nontrivial regions, we propose a generalized 1D method .
\begin{equation}
\nu_{0,\pm\frac{\omega}{2} }\!=\!|\mathrm{mod}(\sum_{n=1}^{n=i}\phi_{Zak}^{n},2)|.
\end{equation}
where $i$ denotes the $i$th bands right below quasi energy $\epsilon\!=\!0,\pm\frac{\omega}{2}$. When $\nu_{0,\pm\frac{\omega}{2}}=1$, there is  supposed to be Majorana edge modes  pinned at $0,\pm\frac{\omega}{2}$. While $\nu_{0,\pm\frac{\omega}{2}}=0$ predicts topological trivial cases. To clarify this, we take Kitaev chain with multiple time periodic driving as an example. When $r_0=1.2$ (other parameters are fixed the same with that in Fig. 1(d)), the Zak phase~\cite{see supplemetary} of Floquet bands is $\mathrm{[-1,-1,-1,-1,-1,-1,-1,-1]}$, hence, $\nu_0=1,\nu_{\pm\frac{\omega}{2}}=0$, there are only MZMs. For $r_0=3$, the Zak phase of the Floquet is $\mathrm{[-1,0,0,0,0,0,0,0]}$, thus  $\nu_0=1,\nu_{\pm\frac{\omega}{2}}=1$. There exist both MZMs and MPMs. Above results coincide with Fig. 1(d).

An alternative way to identify topological properties in driven systems works via the time independent effective Hamiltonian, $U(k,T)\!=\!e^{-i H_{eff}(k) T}$, which is not always applicable unless the eigenvalues of phase bands $\varphi(k,t)$, $u(k,t)\!=\!e^{-i\varphi(k,t)}$, have no degenerate points for $t\neq 0$~\cite{Nathan-NJP-2015}, which ensures the capability of distorting to flat bands. Fortunately, this is assured for Kitaev chain with multiple time periodic driving at high frequency limit. For example, Figs. 3(a)(b) depict the phase band $\varphi(k,t)$ and bulk spectrum of $U(k,T)$ with driving parameters: $r_0=2,r_1=2,\Delta_0=0.5,\Delta_1=2,\omega=10,\mu=2$. Non degenerate points are discovered with nonvanished $t$.
\begin{figure}[t]
 \centering
\includegraphics[width=8.5cm,height=4cm]{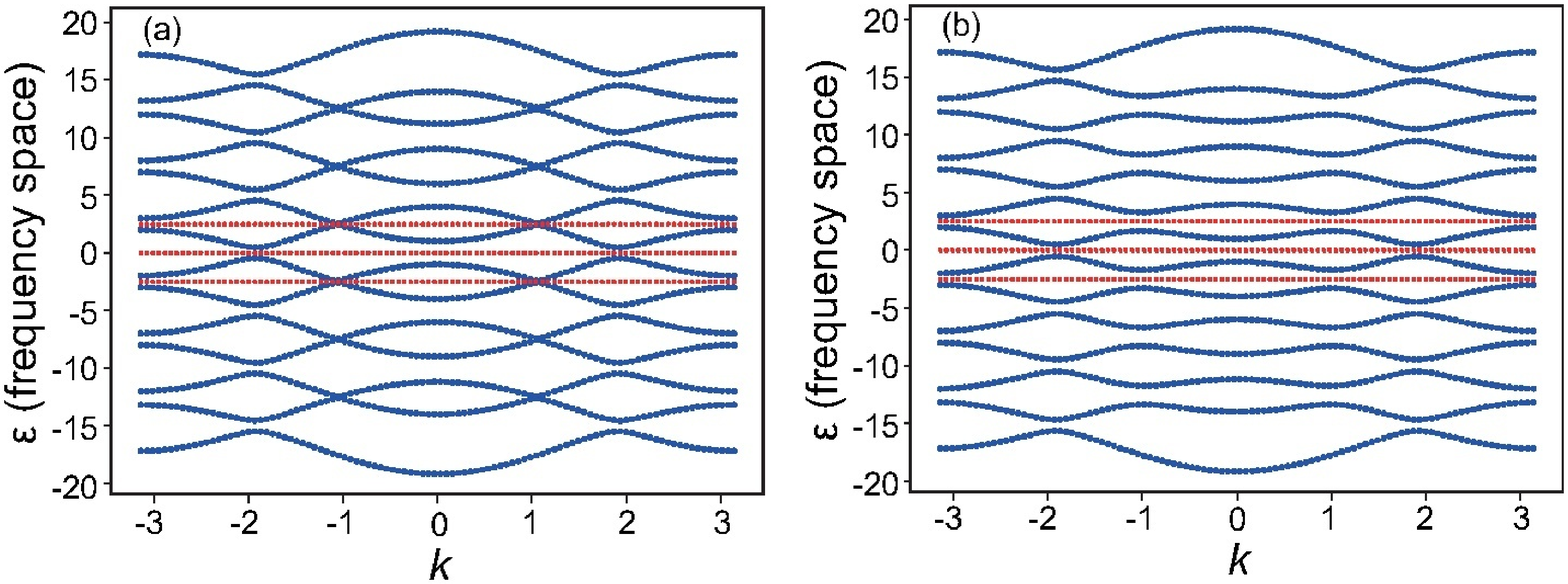}\hspace{3.0cm}
\caption{\label{fig:floquet frequency spectrum}(Color online) Floquet quasi energy spectrum in frequency space with truncation at $m=3$, where $r_0=3,\Delta_0=0.5, \omega=5, \mu=2$. Red dashed line denotes quasi energy $0,\pm\frac{\omega}{2}$. (a) Floquet bands with single time periodic hopping ($r_1=2$)  cross at $\epsilon=\pm \frac{\omega}{2}$, and avoid crossing at $\epsilon=0$. (b) Floquet bands with time periodic hopping ($r_1=2$)  and pairing ($\Delta_1=2$)  driving avoid crossing at $\epsilon=0,\pm\frac{\omega}{2}$.}
\end{figure}
Before proceeding, we would like to change to the rotating frame to obtain a simplified time independent effective Hamiltonian. First, we apply an unitary transformation to the Floquet state $\widetilde{\Phi}(t)\!=\!S^{\dagger}(t)\Phi(t)$. Floquet Schrodinger equation in the rotating frame is $[\widetilde{H}(t)-i\partial_t]\widetilde{\Phi}(t)\!=\!\epsilon \widetilde{\Phi}(t)$, where $\widetilde{H}\!=\!S^{\dagger}H(t)S-iS^{\dagger}\dot{S}$. It is exhibited that in both the original and the rotating frame, Floquet quasi energy is the same, giving rise to unchanged Floquet operator $U(T)\!=\!e^{-i\epsilon T}$, which means the time independent effective Hamiltonian built from either frame shall be equivalent in describing edge modes.

To obtain more concrete results, we resort to the high frequency limits, at which the time independent effective Hamiltonian can be analytically resolved. With periodical boundary condition, the bulk Hamiltonian of Eq. (1) in k space is $H(t)\!=\!\sum_{k}\psi_k^{\dagger}H_k(t)\psi_k$, where $ \psi_k^{\dagger} \!=\!(c_k^{\dagger},c_{-k}), H_k(t)\!=\!-(r(t)\cos{k}\!+\!\frac{\mu_0}{2})\sigma_k^{z}\!+\!\Delta(t)\sin{k}\sigma_k^{y}$. Choosing the rotating frame transformation $S(t)\!=\!e^{i\theta(t)\cos{k}\sigma_k^{z}}, \theta(t)\!=\!\frac{r_1}{\omega}\sin(\omega t)$,  the transformed Hamiltonian can be described by
\begin{equation}
\begin{aligned}
\widetilde{H}(k,t)\!=\!&-(\frac{\mu_0}{2}+r_0\cos{k})\sigma_k^{z}-i\Delta(t)\sin{k}e^{-2i\theta(t)\cos{ k}}\sigma_k^{+}\\
&\!+\!i\Delta(t)\sin{k}e^{2i\theta(t)\cos{k}}\sigma_k^{-}.
\end{aligned}
\end{equation}
It is transparent that the modified Hamiltonian in rotating frame shares the same periodicity as the original one. Applying the Fourier transformation defines $\widetilde{H}_p(k)\!=\!\frac{1}{T}\int_0^T e^{-ip\omega t } \widetilde{H}(k,t)dt$. The time independent effective Hamiltonian can be expanded  via Magus expansion~\cite{Blane-prep-2009}:
\begin{equation}
\begin{aligned}
\widetilde{H}_{eff}(k)=&\widetilde{H}_0(k)+\frac{1}{\omega}[\widetilde{H}_0(k),\widetilde{H}_1(k)]-\frac{1}{\omega}[\widetilde{H}_0(k),\widetilde{H}_{-1}(k)]\\
&-\frac{1}{\omega}[\widetilde{H}_{-1}(k),\widetilde{H}_{1}(k)]+\ldots
\end{aligned}
\end{equation}
To get analytical results, above series should be converged within finite terms  which requires $\int_0^{T}|| \widetilde{H}(k,t)|| dt\!<\!\pi$. In our driving scenario, this is equivalent to $\omega\!>\!2r_0+\mu$. Given all above, the $\widetilde{H}_{p}(k)$ can be denoted by
\begin{equation}
\begin{aligned}
&\widetilde{H}_{p}(k)\!=\!-[r_0\cos{k}\!+\!\frac{\mu_0}{2}]\delta_{p,0}\sigma_k^{z}\!-\![i\Delta_0\sin{k}J_{-p}(\frac{2r_1}{\omega}\cos{k})\\
&+i\frac{\Delta_1}{2}\sin{k}e^{i\phi}J_{-(p-1)}(\frac{2r_1}{\omega}\cos{k})\\
&+i\frac{\Delta_1}{2}\sin{k}e^{-i\phi}J_{-(p+1)}(\frac{2r_1}{\omega}\cos{k})]\sigma_k^{+}\\
&\!+\![i\Delta_0\sin{k}J_{p}(\frac{2r_1}{\omega}\cos{k})\!+\!i\frac{\Delta_1}{2}\sin{k}e^{i\phi}J_{(p-1)}(\frac{2r_1}{\omega}\cos{k})\\
&+i\frac{\Delta_1}{2}\sin{k}e^{-i\phi}J_{(p+1)}(\frac{2r_1}{\omega}\cos{k})]\sigma_k^{-},
\end{aligned}
\end{equation}
where $J_p$ is the $p$th order Bessel function. At high frequency limits, $\widetilde{H}_0(k)$ is the dominant term, and meanwhile notice that the Bessel function decays rapidly in its order.  Hence, the time independent effective Hamiltonian can be approximated as
\begin{equation}
\begin{aligned}
\widetilde{H}_{eff}(k)\!=\!&-[\frac{\mu_0}{2}\!\!+\!\!r_0\cos{k}]\sigma_k^{z}+[\Delta_0\sin{k}J_0(\frac{2r_1}{\omega}\cos{k})]\sigma_k^{y}\\
&+[\Delta_1\sin{k}\sin{\phi}J_1(\frac{2r_1}{\omega}\cos{k})]\sigma_k^{x}.
\end{aligned}
\end{equation}
\begin{figure}[t]
 \centering
\includegraphics[width=8.5cm,height=4cm]{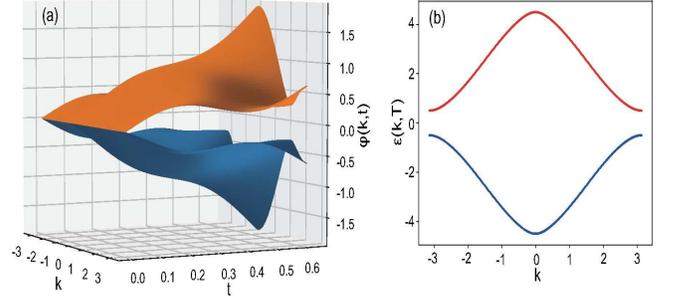}\hspace{3.0cm}
\caption{\label{fig:phaseband}(Color online) Kitaev chain with time periodic hopping and pairing where $r_0=2,r_1=2, \Delta_0=0.5,\Delta_1=2, \omega=10,\mu=2$. (a) The phase band $\varphi(k,t)$ of the driving process. No degeneracies in quasi energy are discovered for $t\neq0$  which ensures the capability of distorting to flat bands. (b) The bulk spectrum of Floquet operator $U(k,T)$.}
\end{figure}
In comparison with the static case, the time independent effective Hamiltonian of  Kitaev chain with multiple driving is composed of terms proportional pauli matrix $\sigma_k^{x,y,z}$. From a progressive view, we set $\phi\!=\!n\pi$ at present (or the time periodic pairing $\Delta_1\!=\!0$).
The terms proportional to  $\sigma_k^{x}$ are vanished, and the differences to the static Kitaev chain lie in the renormalized pairing $\Delta_{eff}(k)\!\!=\!\!\Delta_0 J_0(\frac{2r_1}{\omega}\cos{k})$. Considering the unitary Majorana-fermion transformation $
   U\!=\!\frac{1}{\sqrt{2}} \begin{bmatrix}
   1& i\\
   1& -i\\
   \end{bmatrix}$
and the transformed Hamiltonian is denoted by
\begin{equation}
\begin{aligned}
\widetilde{H'}_{eff}(k)=& U^{\dagger}HU=\begin{bmatrix}
 0& h(k) \\
h^{\ast}(k) & 0\\
 \end{bmatrix}
\end{aligned}
\end{equation}
in which the off diagonal entry is $h(k)\!=\!-i(\frac{\mu_0}{2}\!+\!r_0\cos{k})\!-\!\Delta_0\sin{k}J_0(\frac{2r_1}{\omega}\cos{k})$. The winding number $W$ can be defined via the chiral index: $W\!=\!\int_{-\pi}^{\pi} \frac{dk}{2\pi i} \partial_k \ln h(k)$.
\begin{figure}[t]
 \centering
\includegraphics[width=8.5cm, height=7cm]{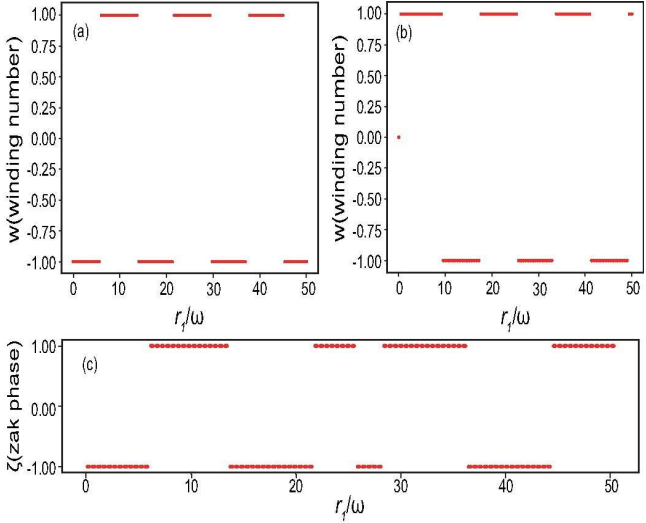}\hspace{3.0cm}
\caption{\label{fig:index}(Color online) (a),(b) the winding number $W$ varies discretely as a function of ratio $\frac{r_1}{\omega}$, where $r_0=5$, $\mu=2$. The relative phase $\phi=n\pi, n\in Z $, and $\Delta_0=5$ for (a).  The relative phase $\phi=\frac{\pi}{2}$ and $\Delta_0=0, \Delta_1=5$  for (b). (c) When the relative phase $\phi=\frac{\pi}{3},\Delta_0=5$, the time independent effective Hamiltonian is composed of terms proportional to $\sigma_k^{x,y,z}$. Zak phase varies discretely as a function of $\frac{r_1}{\omega}$ with $r_0=5,\Delta_1=5,\mu=2$.}
\end{figure}
Numerical results reveal that the driven Kitaev chain at high frequency limit has the same topological phase transition boundary as the static one ($2r_0\!>\!\mu$). However, the sign of winding number not only relies on the sign of pairing but also the ratio $\frac{r_1}{\omega}$, which is depicted in Fig. 4(a). According to the bulk edge correspondence, $W\!=\!1$ implies that there are only one pair of Majorana edge modes, which features the characteristics of high frequency limit.

We then continue to  the case $\phi\neq n\pi$. First, we set $r_1=0$.  Noticing that $J _1(0)\!=\!0, J_0(0)\!=\!1$, the effective Hamiltonian in Eq. (7) transform backs to the static case, which means that at high frequency limits, only time periodic pairing has no impacts on topological properties to distinguish itself from a non driven Kitaev chain.  Now we proceed to the case $ r_1\!\neq\!0,\Delta_0\!=\!0$,  the  $\widetilde{H}_{eff}(k)$ only contains terms proportional to $\sigma_k^{x,z}$. The system still possesses the particle-hole  symmetry, $\Xi \widetilde{H}_{eff}(k)\Xi^{-1}\!=\!-\widetilde{H}_{eff}(-k)$, $\Xi\!=\!\sigma^{x}$, supporting the existence of Majorana edge modes. To seek for topological index, we utilize similar operations as Eqs. (8),(9). Given the unitary transformation:
$
   U'\!=\!\frac{1}{\sqrt{2}} \begin{bmatrix}
   1& i\\
   i& 1\\
   \end{bmatrix}
$.
The time independent effective Hamiltonian can be transformed to forms with only off-diagonal entry: $h'(k)=-i(\frac{\mu_0}{2}+r_0\cos{k})+\Delta_1\sin{k}J_1(\frac{2r_1}{\omega}\cos{k}) $. The winding number also varies discretely between $-1,1$ as a function of ratio $\frac{r_1}{\omega}$, see Fig. 4(b).  Need to mention that when $r_1=0,\Delta_0=0$, effective Hamiltonian in Eq. (7) only contains term proportional to $\sigma_k^{z}$, and the system is topological trivial. Finally, we release all the restrictions and $r_1\!\neq\!0, \Delta_0\!\neq\!0$. Terms proportional to $\sigma_k^{x,y,z}$ all exist, and the time independent effective Hamiltonian shall be indexed by Zak phase: $\zeta=\frac{1}{2\pi i}\int dk\langle u(k)|\partial_k u(k)\rangle $, where $|u(k)\rangle$ is the eigenstate of the Hamiltonian in Eq. (7). Fig. 4(c) depicts the Zak phase as a function of $\frac{r_1}{\omega}$. In contrast to the previous case, the Zak phase can not be predicted readily with zeros of Bessel function. Meanwhile, we shall notice that when $k=0,\pi$, $\widetilde{H}_{eff}(k)$ in Eq. (7) has only $\sigma_k^{z}$ components, of which the coefficient is defined respectively as $f_0$ (for $k\!=\!0$), $f_\pi$ (for $k\!=\!\pi$). A $Z_2$ index can be denoted as $\mathit{Q}\!=\!sgn(f_0f_\pi)$ to classify topological phases.  $\mathit{Q}\!=\!-1$ predicts topological nontrivial regions, which is the same as a static Kitaev chain.

In conclusion, we have discussed the Kitaev chain with multiple time periodic driving. Our results demonstrate that the multiple driving will introduce diverse avoidance of bands crossing, leading to extraordinary fruitful topological phase transitions. We demonstrate how to identify MPMs, MZMs via the Zak phase of Floquet bands, which can be generalized to other one dimensional driven systems. Differences between single and multiple time periodic driving are manifested analytically at high frequency limit.  Our findings enrich the physics of locating Majorana edge modes in driven system, and will intermediately benefit to the study of quantum computation.

\begin{acknowledgments}
 We are very grateful to M. S. Rudner, E. Zhao, Ahmet Keles, Daniel J. Yates, Monica Benito, for many helpful discussions. This work was supported by the National Key R and D Program of China under grants No. 2016YFA0301500, NSFC under grants Nos.  1835013, 11728407,  the Strategic Priority Research Program of the Chinese Academy of Sciences under grants Nos. XDB01020300, XDB21030300.
\end{acknowledgments}

\end{document}